\documentclass{aa}
\usepackage[varg]{txfonts}
\usepackage[utf8]{inputenc}

\usepackage{graphicx}
\usepackage{enumitem}
\usepackage{txfonts}
\usepackage{textgreek}
\begin{document}

   \title{Machine learning for understanding pulsating stars I:}

   \subtitle{the non-linear phenomenon in $\delta$ Scuti stars}

   \titlerunning{Machine learning for understanding pulsating stars I}

   \author{J.R. Rodon,\inst{1}
          \ J. Pascual-Granado,\inst{1}
          \ M. Lares-Martiz, \inst{2}
          \ M. Rodríguez Sánchez, \inst{3}
          \ C. Roche \inst{4}
          }

    \authorrunning{Rodon et al}

   \institute{
   Instituto de Astrofísica de Andalucía (CSIC). Glorieta de la Astronomía s/n. 18008, Granada, Spain
   \and Embry-Riddle Aeronautical University (ERAU), 1 Aerospace Blvd., Daytona Beach, FL 32114, USA
   \and Departament d’Astronomia i Astrofísica, Universitat de Valéncia, C/ Dr. Moliner, 50, Burjassot, 46100, Valencia, Spain
   \and The Exploration Company, Behringstraße 6, 82152 Planegg, Bavaria, Germany
   }



 
  \abstract
   {$\delta$ Scuti stars are pulsating variable stars that exhibit both radial and non-radial pulsations, making them key objects for understanding stellar evolution and internal structures. The current classification of $\delta$ Scuti stars into High-Amplitude $\delta$ Scuti (HADS) and Low-Amplitude $\delta$ Scuti (LADS) stars is based solely on the peak-to-peak amplitude of their light curves (> 0.3 mag). Nevertheless, this classification may not fully capture the complexity of their pulsation mechanisms and non-linear effects, leading to possible misclassifications and ambiguities.}
   {This investigation aims to challenge the existing classification of $\delta$ Scuti stars according to amplitude, employing the exploration of frequency domain characteristics and non-linear mechanisms in order to identify intrinsic subgroups. The objective is to get a more profound understanding of the properties of the $\delta$ Scuti stars.
   }
   {We use machine learning clustering techniques, specifically hierarchical clustering (HC) with Ward's linkage, to analyze a sample of 142 $\delta$ Scuti stars observed by space telescopes such as CoRoT, Kepler, and TESS. The light curves were processed and analyzed using the Best Parent Method (BPM), which extracts non-linearities and identifies parent and child frequencies. We focus on frequency-domain features, including fundamental and overtone modes, as well as non-linear features such as harmonic, sums, and subtraction frequencies of the fundamental modes, to uncover intrinsic subgroups within $\delta$ Scuti stars.}
   {The results of the clustering process indicate that the present amplitude-based classification (HADS/LADS) exhibits partial alignment with the clusters identified by using features from the frequency-domain. However, the study identified additional subgroups, suggesting a greater variety of nonlinear effects that are not captured by the amplitude alone. It highlights the importance of non-linear features, such as the number of subtraction combinations, which may be indicative of resonance effects or other internal stellar mechanisms. 
   }
   {}

   \keywords{$\delta$ Scuti stars, machine learning, clustering, non-linear pulsations.}
   \maketitle
   \nolinenumbers
%

\section{Introduction}

$\delta$ Scuti stars are a class of pulsating variable stars that exhibit both radial and non-radial pulsations \citep{aerts2021}, making them valuable objects for studying stellar evolution and internal structures \citep{kurtz2022asteroseismology}. These stars are located in the lower part of the instability strip on the Hertzsprung-Russell (HR) diagram, with spectral types ranging from A2 to F2 and effective temperatures between 6300 and 8900 K \citep{breger1998period}. Their pulsations are driven by the $\kappa$-mechanism, in the He II partial ionization layers, leading to periodic changes in opacity and temperature \citep{cox1980}. This mechanism causes them to expand and contract, producing observable variations in their luminosity and effective temperature.

The study of $\delta$ Scuti stars has evolved significantly with the advent of space-based observatories such as CoRoT \citep{auvergne2009corot}, Kepler, and TESS \citep{ricker2015transiting}, which provide high-precision photometric data. These missions have revolutionized our understanding of pulsating stars by enabling continuous monitoring and the detection of subtle brightness variations that were previously unobservable from ground-based telescopes. The high-quality data from these missions have revealed complex pulsation spectra \citep{handler2011hybrid}, including both pressure (p) and gravity (g) modes, as well as non-linear effects such as harmonics, sums, and subtraction frequencies of these modes, namely non-linearities \citep{balona2012combination, balona2016combination}. These non-linearities may arise from interactions among different pulsation modes and from interactions between pulsation modes and the stellar medium, referred to as pulsation–pulsation and pulsation–medium interactions \citep{bowman2018characterizing, Lares-Martiz2022}. Consequently, they can serve as potential probes of the internal structure and dynamical processes operating in these stars.

Despite these advances, the internal classification of $\delta$ Scuti stars remains largely based on the peak-to-peak amplitude of their light curves \citep{mcnamara2000high}, dividing them into High-Amplitude $\delta$ Scuti (HADS) and Low-Amplitude $\delta$ Scuti (LADS). This classification, established over two decades ago, uses a threshold of 0.3 magnitudes in peak-to-peak amplitude to distinguish between the two groups. However, this approach has been questioned for its simplicity and the potential to overlook underlying physical mechanisms \citep{aerts2021}. For instance, the light curve amplitude results from the combined effects of pulsation mechanisms and non-linear interactions within the star, making it a potentially misleading indicator of the star's internal properties.

This study aims to assess the limitations of current amplitude-based classification by applying machine-learning clustering techniques to a dataset of 142 $\delta$ Scuti stars. The study focuses on frequency-domain features, including fundamental and overtone modes, as well as non-linearities, to identify intrinsic subgroups within $\delta$ Scuti stars. The objective of this study is to provide a more comprehensive understanding of the pulsation mechanisms and internal structure of these stars. The ultimate goal is to contribute to the broader field of stellar astrophysics by proposing a more natural classification system that better reflects the physical properties and mechanisms driving the pulsations in $\delta$ Scuti stars.

The paper begins with an introduction that establishes the importance of δ Scuti stars and the limitations of the current amplitude-based classification. The methodology is detailed in Section 2, covering data preprocessing, feature selection, and outlier detection. Section 3 is dedicated to the application and evaluation of various clustering techniques, ultimately selecting hierarchical clustering with Ward's linkage, and presents the clustering results and their discussion in subsections. The paper concludes with Section 4, which summarizes the findings, discusses their implications, and outlines several promising directions for future research, followed by a comprehensive list of references.


\section{Exploratory Data Analysis and Feature Selection}

This section details the preprocessing and feature selection procedures applied to a light curves to prepare the data for clustering analysis. 

The sample analysed in this study comprises light curves of 142 $\delta$ Scuti stars obtained from high-precision photometric observations by the CoRoT, Kepler (short-cadence mode), and TESS space telescopes. From these light curves, a set of features was extracted for the clustering analysis, including fundamental and overtone mode frequencies, amplitudes, phases, and indicators of non-linear behaviour. The goal is to identify the most informative features that capture the pulsation mechanisms and non-linear effects in $\delta$ Scuti stars while minimizing noise and bias.

The pulsation parameters were determined using the BPM algorithm \citep{Mariel2021}, a self-consistent approach designed to identify and remove combination frequencies in pulsating star light curves. First, the user selects candidate parent frequencies as initial seeds. In this study, these seeds were selected by searching for frequency pairs that satisfy the Stellingwerf relations. When no such pair was found, the frequencies with the highest amplitudes in the power spectrum were adopted, or only the dominant frequency in the case of monoperiodic stars. The algorithm then explores the parent-frequency space by fitting non-linear models (e.g., Volterra expansions) and computing the variance of the residuals, iteratively refining the parent frequencies with progressively smaller frequency steps until a minimum variance (“best” parents) is reached. Finally, these optimal parent frequencies are fitted simultaneously with all statistically significant combination (“child”) frequencies in a least-squares framework, yielding residuals largely free of non-linearities.

The refined fundamental-mode parameters—frequency (\(f_1\)), amplitude (\(A_1\)), and phase (\(P_1\))—together with, when present, the first overtone frequency (\(f_2\)), amplitude (\(A_2\)), and phase (\(P_2\)), were used as input features for the clustering analysis. For monoperiodic stars, \(f_2\) and \(A_2\) were set to zero to encode the absence of a secondary mode. Non-linear features, such as the number of harmonics and combination frequencies (both additive and subtractive), were also included. These features were chosen to represent the degree of the star’s non-linear behaviour.




The goal is to identify the most informative features that capture pulsation mechanisms and non-linear effects in $\delta$ Scuti stars while minimizing noise and bias.

\subsection{Preprocessing of Raw Data Features}
The raw data were preprocessed to ensure consistency and improve statistical properties. Features such as fundamental mode frequency (here we used the notation \(f_1\)), amplitude (\(A_1\)), and phase (\(P_1\)), as well as first overtone mode frequency (\(f_2\)) and amplitude (\(A_2\)), were selected for analysis. Monoperiodic stars, which lack an overtone mode, were encoded with zero values for \(f_2\) and \(A_2\), as a frequency of 0 Hz and zero amplitude represent the absence of a signal in the overtone mode. Preliminary studies have not found P2 to have a significant influence on the results, and we therefore prefer not to include it, as any value would introduce artificial correlations. This prioritizes a clean and meaningful dataset. This avoided injecting artificial patterns and forced the clustering algorithm to focus on more robust features. PSD values were transformed into amplitudes to reduce variance in the clustering and improve performance.

Non-linear features, such as the number of harmonics and combination frequencies (both additive and subtractive), were also included to capture the complexity of the stars' pulsation spectra. These features were chosen to represent the degree of non-linear effects rather than the specific characteristics of individual modes. The final feature set comprised nine variables, balancing the need for informative features with the constraints of the dataset size.

\subsection{Outlier Detection}
Outliers were identified using a combination of statistical methods, distance-based techniques, and density-based clustering. Initial outlier detection methods, such as z-scores and interquartile ranges,  were tested but failed due to the non-Gaussian distributions of the data. To address this, logarithmic transformations were applied to features with exponential-like distributions, making them more suitable for statistical outlier detection. However, these methods still struggled to reliably distinguish between genuine outliers and natural variability.

Distance-based methods, such as K-means clustering, were also tested but were found to be sensitive to the presence of extreme values and the assumption of spherical clusters. Density-based methods, particularly DBSCAN \citep{breunig2000lof} (Density-Based Spatial Clustering of Applications with Noise), were more robust for handling skewed distributions and extreme values. DBSCAN identifies outliers as points that do not belong to any dense cluster, but its performance depends heavily on the choice of parameters $\epsilon$ (radius around each data point) and minPts (the minimum number of data points that must be within the $\epsilon$-radius of a point for that point to be considered a core point), which can be challenging to tune.

Ultimately, outlier detection was integrated into the clustering process, with hierarchical clustering (HC) using Ward's linkage proving to be the most reliable method. HC is robust to outliers and can handle non-linear distributions, making it well-suited for the dataset. Outliers were treated as potential natural variations or misclassified stars, rather than errors, and were included in the clustering analysis to avoid losing valuable information.

\subsection{Feature Selection Techniques}
Several feature selection techniques were tested to validate the chosen features and reduce dimensionality. Dimensionality reduction methods, such as Principal Component Analysis (PCA), Multidimensional Scaling \citep {davison2000multidimensional} (MDS), and t-SNE \citep{van2008visualizing}, were evaluated. PCA, which linearly combines variables to capture maximum variance, was limited by its assumption of linear relationships and sensitivity to outliers. While PCA provided some insights, it did not significantly improve clustering performance and resulted in the loss of interpretability.

MDS, which preserves pairwise distances between data points, was more suitable for capturing global structures but still struggled with non-linear relationships. t-SNE, by contrast, prioritizes local structure and non-linear relationships, making it ideal for visualizing clustering patterns in high-dimensional data. Although t-SNE was not used for feature selection due to its lack of interpretability, it was employed to validate the clustering results by mapping high-dimensional data into a lower-dimensional space for visualization.

The final feature set included nine parameters (see Table \ref{tab:features}), selected based on their physical relevance and ability to capture the pulsation mechanisms and non-linear effects in $\delta$ Scuti stars. This set provided a balance between informative features and the constraints of the dataset size, ensuring that the clustering analysis would be both meaningful and computationally feasible.

\begin{table}[]
    \centering
    \begin{tabular}{lcr}
Description & Parameter & Units\\
\hline
Fundamental mode frequency & f1 & $d^{-1}$\\
Fundamental mode amplitude & A1 & $(e^{-}/s)^2$ \\
Fundamental mode phase & P1 & rad \\
Overtone mode frequency & f2  & $d^{-1}$\\
Overtone mode amplitude & A2 & $(e^{-}/s)^2$ \\
Nº of fundamental mode harmonics & Harm1 & -\\
Nº of overtone mode harmonics & Harm2 & -\\
Nº of addition combinations & AddComb & -\\
Nº of subtraction combinations & SubComb & -\\
\hline
\vspace{0.3cm}
\end{tabular}
    \caption{Description of the nine parameters included in the final feature set.}
    \label{tab:features}
\end{table}

This section highlights the importance of careful preprocessing and feature selection in preparing the dataset for clustering analysis. The chosen features and techniques aim to capture the underlying physical mechanisms of $\delta$ Scuti stars while minimizing noise and bias, setting the stage for the application of clustering algorithms in the subsequent sections.


\section{Data-Driven Group Discovery in $\delta$ Scuti Stars Using Clustering Methods}


This section explores the application of clustering techniques to identify intrinsic subgroups within $\delta$ Scuti stars based on Fourier parameters. The goal is to uncover patterns that may not be captured by the traditional amplitude-based classification (HADS/LADS) and to provide a more nuanced understanding of the pulsation mechanisms and internal structures of these stars.

\subsection{Roadmap of the Selection of Clustering Techniques}

The selection of clustering techniques is a critical step in this study, as the choice of method directly impacts the ability to uncover meaningful subgroups within the $\delta$ Scuti stars dataset. Given the complexity of the dataset, which includes non-linear relationships, skewed distributions, and potential outliers, several clustering methods were evaluated to identify the most suitable approach. The goal was to select a method that could handle the dataset's characteristics while providing interpretable and reliable results.

\subsubsection{Centroid-Based Methods}
Centroid-based clustering methods, such as K-means, were initially considered due to their computational efficiency and conceptual simplicity. These methods group data points based on their proximity to randomly initialized centroids, which are iteratively updated to minimize the distance between points and their assigned centroids.

\begin{itemize}
    \item \textbf{Strengths:} K-means is computationally efficient and works well for datasets with spherical clusters of similar size and density.
    \item \textbf{Limitations:} K-means is highly sensitive to the initialization of centroids and the choice of the number of clusters (\(k\)). It assumes that clusters are spherical and struggles with non-convex or irregularly shaped clusters. Additionally, K-means is not robust to outliers, as a single outlier can significantly distort the centroid positions and cluster boundaries.
    \item \textbf{Results:} When applied to the $\delta$ Scuti dataset, K-means produced clusters that were heavily influenced by extreme values, leading to unreliable results. The method demanded extensive parameter tuning, and the results exhibited low consistency across multiple runs.
\end{itemize}

\subsubsection{Distribution-Based Methods}
Distribution-based clustering methods, such as Gaussian Mixture Models (GMM), were also evaluated. These methods assume that the data is generated from a mixture of probability distributions, typically Gaussian distributions, and use the Expectation-Maximization (EM) algorithm to estimate the parameters of these distributions.

\begin{itemize}
    \item \textbf{Strengths:} GMM allows for more flexible cluster shapes compared to K-means, as it can model ellipsoidal clusters of varying sizes and orientations.
    \item \textbf{Limitations:} GMM assumes that the data follows a mixture of Gaussian distributions, which may not hold for features with skewed or non-Gaussian distributions. Additionally, GMM is sensitive to the initialization of parameters and can struggle with high-dimensional data.
    \item \textbf{Results:} When applied to the $\delta$ Scuti dataset, GMM produced clusters that did not align well with the underlying structure of the data. The method struggled to capture the complex relationships between features, particularly for non-linear and zero-inflated distributions.
\end{itemize}

\subsubsection{Density-Based Methods}
Density-based clustering methods, such as DBSCAN (Density-Based Spatial Clustering of Applications with Noise), were evaluated for their ability to handle arbitrary-shaped clusters and robustness to outliers. DBSCAN groups data points based on their density, identifying dense regions separated by areas of lower density.

\begin{itemize}
    \item \textbf{Strengths:} DBSCAN does not require the number of clusters to be specified in advance and can identify clusters of arbitrary shapes. It is also robust to outliers, as points that do not belong to any dense region are labeled as noise.
    \item \textbf{Limitations:} DBSCAN requires careful tuning of its parameters, particularly the neighborhood radius (\(\epsilon\)) and the minimum number of points (\(minPts\)) required to form a cluster. The method can struggle with datasets that have varying densities, as a single set of parameters may not capture all clusters accurately.
    \item \textbf{Results:} When applied to the $\delta$ Scuti dataset, DBSCAN identified some clusters but struggled with the varying densities and non-linear relationships in the data. The method required extensive parameter tuning, and the results were not consistent across different runs.
\end{itemize}

\subsubsection{Hierarchical Methods}
Hierarchical clustering (HC) methods, particularly agglomerative clustering with Ward's linkage, were ultimately selected for their robustness to skewed data and outliers. HC builds a tree-like structure (dendrogram) that represents the relationships between data points, allowing for the exploration of clusters at different levels of granularity. 

\begin{itemize}
    \item \textbf{Strengths:} HC does not require the number of clusters to be specified in advance and provides a visual representation of the clustering process through the dendrogram. Ward's linkage, which minimizes the variance within clusters, is particularly well-suited for datasets with non-Gaussian distributions and varying cluster sizes.
    \item \textbf{Limitations:} HC can be computationally expensive for large datasets, as it requires calculating pairwise distances between all data points. Additionally, the method is sensitive to the choice of linkage criteria and can struggle with high-dimensional data.
    \item \textbf{Results:} When applied to the $\delta$ Scuti dataset, HC with Ward's linkage produced the most reliable and interpretable results. The method was able to capture the underlying structure of the data, including non-linear relationships and varying densities. The dendrogram provided a clear visualization of the clustering process, allowing for the identification of both main groups and subgroups within the dataset.
\end{itemize}

\begin{figure}[h]
\centering
  \includegraphics[width=0.48 \textwidth]{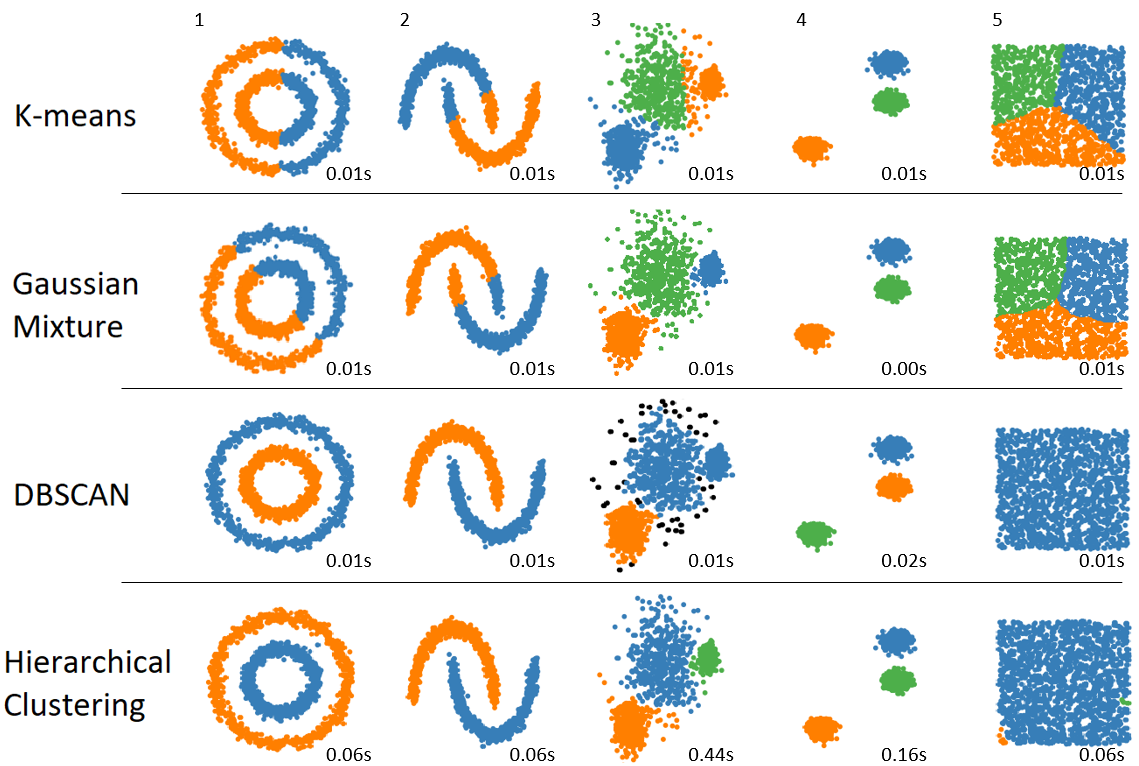}
  \caption{Performance comparison against different types of distributions (columns) of the considered clustering techniques (rows). Image modified from \citep{scikit-learn}.}
  \label{fig:clusters}
\end{figure}

\subsubsection{Discussion}
Fig.~\ref{fig:clusters} compares the performance of various clustering techniques against different types of data distributions. We include it here to provide a visual rationale for our choice of clustering methodology, as it succinctly illustrates the expected behavior of similar techniques across diverse scenarios. 

The evaluation of clustering techniques highlighted the importance of selecting a method that can handle the complex characteristics of the $\delta$ Scuti dataset. While centroid-based and distribution-based methods struggled with non-linear relationships and skewed distributions, hierarchical clustering with Ward's linkage proved to be the most robust and interpretable approach. The selection of HC was further supported by its ability to handle outliers and provide a visual representation of the clustering process, making it well-suited for uncovering intrinsic subgroups within $\delta$ Scuti stars.

This roadmap sets the stage for the application of hierarchical clustering in the subsequent sections, with the goal of identifying meaningful patterns and refining the classification of $\delta$ Scuti stars based on their pulsation mechanisms and non-linear effects.

\subsection{Clustering Results}

This section presents the results of applying hierarchical clustering (HC) with Ward's linkage to the dataset of 142 $\delta$ Scuti stars. The clustering analysis was performed using two main approaches: (1) a baseline clustering based solely on the fundamental mode amplitude, and (2) a more comprehensive clustering using a 9-dimensional feature space that includes both linear and non-linear features. The results are discussed in detail, with a focus on their implications for the classification of $\delta$ Scuti stars.

\subsubsection{Clustering Based on the Fundamental Amplitude}
As a baseline, clustering was performed using only the fundamental mode amplitude (\(A_1\)), which is closely related to the traditional amplitude-based classification (HADS/LADS). The goal was to compare the results of this simple clustering with the more comprehensive approach using multiple features.

\begin{figure*}[h]
\centering
  \includegraphics[width=0.9\textwidth]{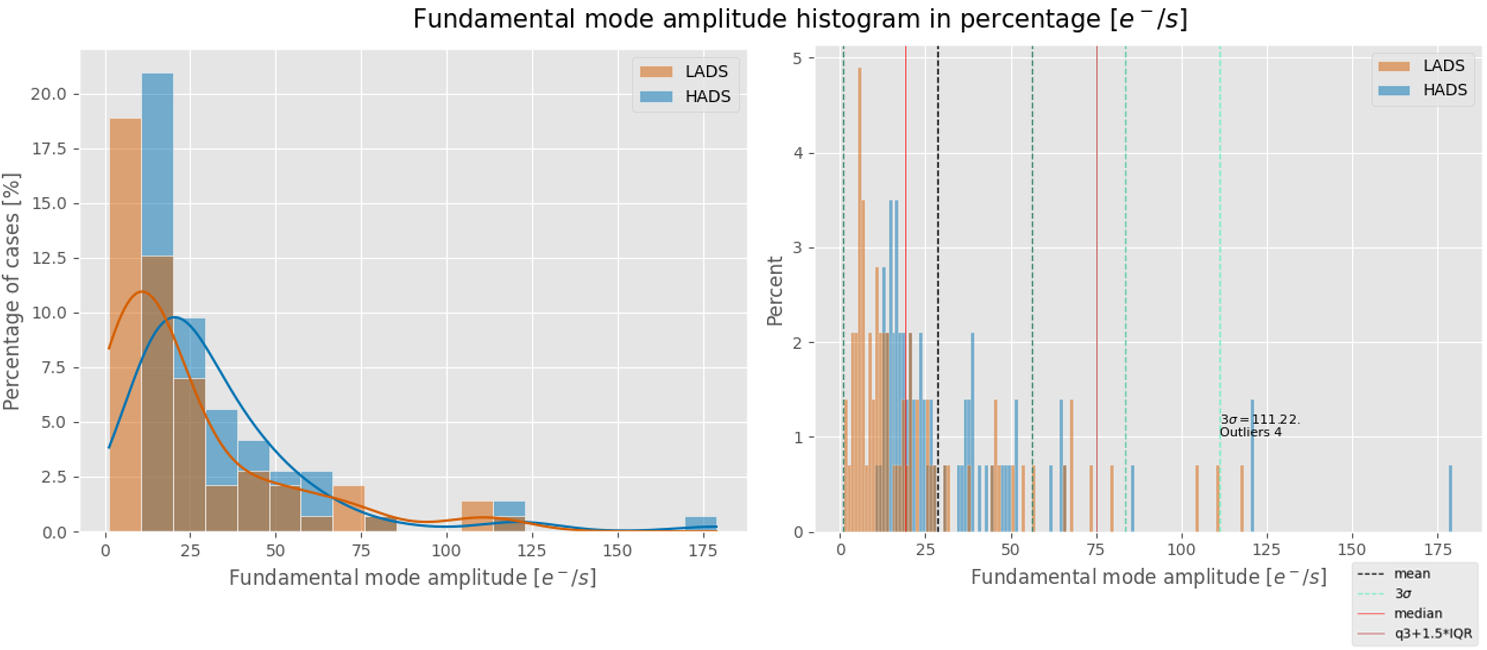}
  \caption{Fundamental amplitude histogram. On the left, depicted with thicker bins to show the tendency. On the right, with unitary bins and statistical measures for distribution analysis.}
  \label{fig:A1}
\end{figure*}

\begin{itemize}
    \item \textbf{Results:} The clustering based on \(A_1\) (see Fig. \ref{fig:A1}) revealed a partial overlap between the distributions of HADS and LADS. While HADS generally exhibited higher amplitudes, some LADS were found in the high-amplitude region, and vice versa. This suggests that the fundamental amplitude alone is not sufficient to fully differentiate between HADS and LADS.
    \item \textbf{Discussion:} The overlap in the distributions of HADS and LADS indicates that the light curve amplitude, which is the system's output, masks the underlying pulsation mechanisms and non-linear effects. This supports the idea that a more comprehensive approach, incorporating multiple features, is needed to accurately classify $\delta$ Scuti stars. 
\end{itemize}

\subsubsection{HC Clustering Results, 9-Frequency-Domain Feature Space}
Hierarchical clustering (HC) was applied to the 9-dimensional feature space, which includes fundamental and overtone mode frequencies, amplitudes, and non-linear features such as harmonics and combination frequencies. The results are presented for two scenarios: clustering into 3 groups and clustering into 6 groups.

We deliberately bypassed a forced 2-cluster solution to avoid presupposing that the HADS/LADS dichotomy is the fundamental ground truth. Testing with 3 and 6 groups allowed the data to reveal its intrinsic structure, whether it confirmed the traditional binary scheme or, as our results indicate, uncovered more complex subgroups and phenomena (like a resonance-driven group) that a simple 2-cluster approach would have obscured.

\begin{figure*}[h]
\centering
  \includegraphics[width=0.7\textwidth]{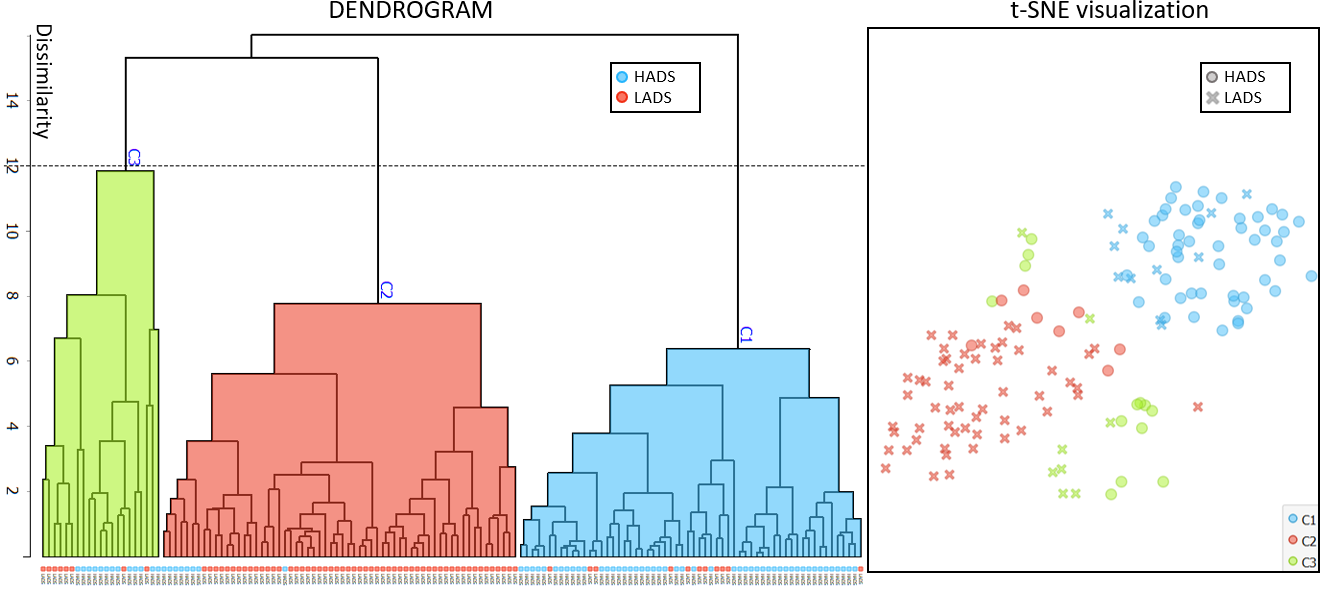}
  \caption{Dendrogram and t-SNE visualisation of the 3 clusters obtained by applying HC with Ward's method to the 142 $\delta$ Sct sample using 9-frequency-domain features.}
  \label{fig:9HC3}
\end{figure*}

\begin{figure*}[h]
\centering
  \includegraphics[width=0.7\textwidth]{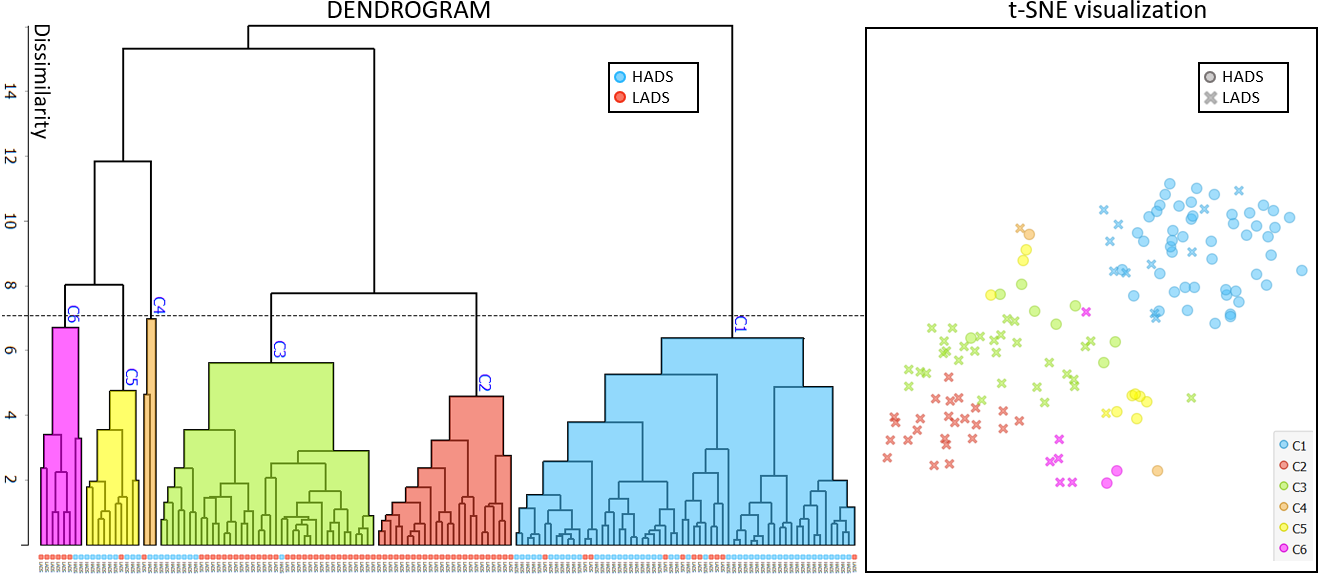}
  \caption{Dendrogram and t-SNE visualisation of the 6 clusters obtained by applying HC with Ward's method to the 142 $\delta Sct$ sample using 9-frequency-domain features.}
  \label{fig:9HC6}
\end{figure*}

\begin{itemize}
    \item \textbf{3 Clusters:} The initial clustering into three groups showed a significant alignment with the traditional HADS/LADS classification. However, a third cluster emerged, characterized by a high number of subtraction combinations, suggesting the presence of resonance effects or other non-linear mechanisms (see Fig.\ref{fig:9HC3}).
    
    \begin{itemize}[label=\textbullet]
        \item \textbf{Cluster C1 (blue):} This cluster primarily consisted of HADS and was characterized by monoperiodic stars with higher amplitudes and fewer non-linear features.
        
        \item \textbf{Cluster C2 (red):} This cluster primarily consisted of LADS and was characterized by multi-periodic stars with lower amplitudes and a moderate number of non-linear features.
        
        \item \textbf{Cluster C3 (green):} This cluster included stars with a high number of subtraction combinations, indicating potential resonance effects. The stars in this cluster did not align neatly with the HADS/LADS classification, suggesting that they represent a distinct subgroup.

    \end{itemize}

    \item \textbf{6 Clusters:} When the number of clusters was increased to six (See Fig.\ref{fig:9HC6}), additional subgroups within the LADS category were identified. These subgroups were characterized by differences in fundamental and overtone frequencies, as well as the number of subtraction combinations, indicating potential variations in pulsation mechanisms or internal structures.
    
    \begin{itemize}[label=\textbullet]
        \item \textbf{Cluster C4:} A subgroup of LADS with lower fundamental and overtone frequencies, potentially indicating stars with different internal structures or evolutionary stages.
        
        \item \textbf{Cluster C5:} A subgroup of LADS with a higher number of subtraction combinations, suggesting the presence of resonance effects or other non-linear mechanisms.
        
        \item \textbf{Cluster C6:} A subgroup of LADS with intermediate characteristics, potentially representing a transition between the other subgroups.
    \end{itemize}
\end{itemize}

\subsubsection{HC Clustering Nonlinear Frequency-Domain Feature Space}
To further explore the role of non-linear effects, clustering was performed using only the non-linear features (harmonics and combination frequencies). The results showed that the non-linear features alone could partially reproduce the HADS/LADS classification, particularly for HADS. This suggests that non-linear mechanisms play a significant role in differentiating HADS from LADS. However, the clustering of LADS required a larger number of clusters, indicating that their classification is more dependent on the natural pulsation modes rather than non-linear effects.

\begin{figure*}[h]
\centering
  \includegraphics[width=0.7\textwidth]{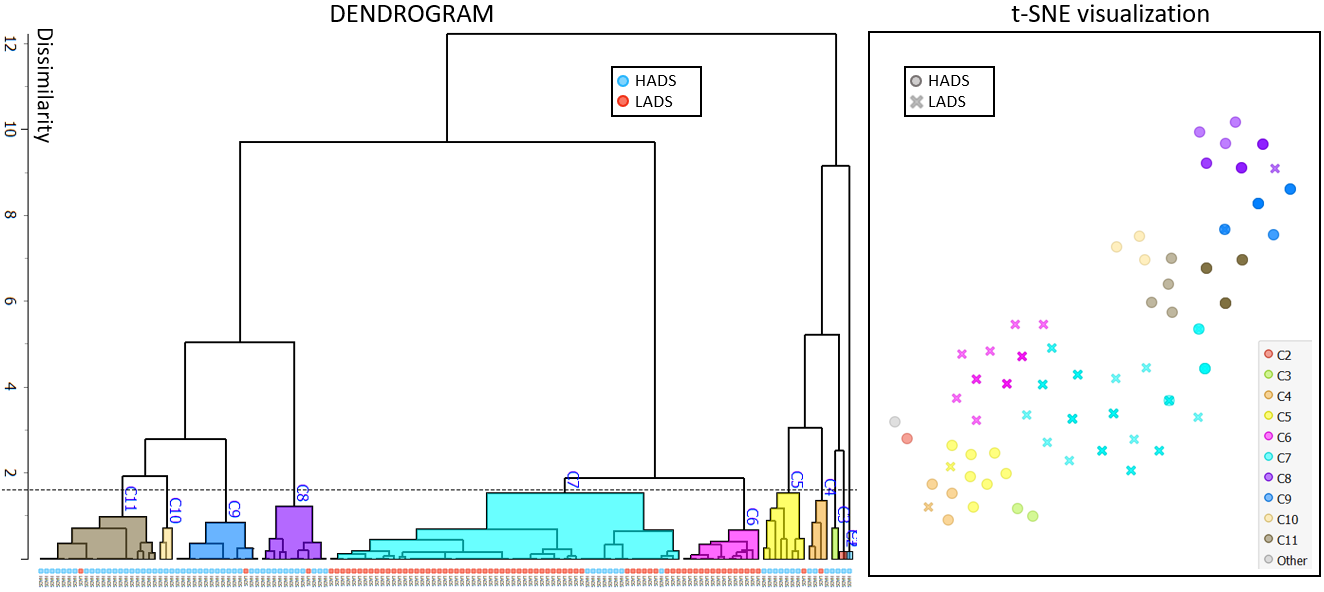}
  \caption{Dendrogram and t-SNE visualisation of the 11 clusters obtained by applying HC with Ward's method to the 142 $\delta$ Sct sample using 3-non-linear frequency-domain features.}
  \label{fig:3HC11}
\end{figure*}

\begin{itemize}
    \item \textbf{3 Clusters:} The clustering into three groups (See Fig. \ref{fig:3HC3}) using only non-linear features revealed a significant alignment with the HADS/LADS classification, particularly for HADS. The third cluster (blue), characterized by a high number of subtraction combinations, again suggested the presence of resonance effects. 

\begin{figure*}[h]
\centering
  \includegraphics[width=0.7\textwidth]{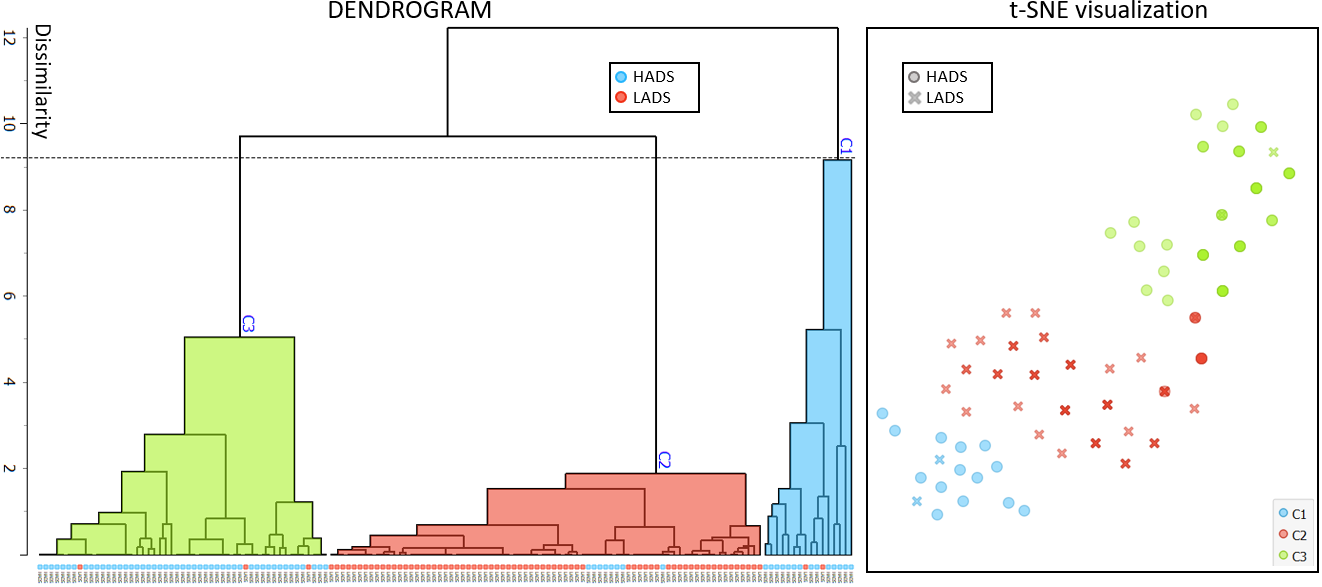}
  \caption{Dendrogram and t-SNE visualisation of the 3 clusters obtained by applying HC with Ward's method to the 142 $\delta$ Sct sample using 3-non-linear frequency-domain features.}
  \label{fig:3HC3}
\end{figure*}
    
    \item \textbf{11 Clusters:} When the number of clusters was increased to eleven (See Fig. \ref{fig:3HC11}), additional subgroups within the LADS category were identified. However, these subgroups did not align well with the clusters identified using the 9-dimensional feature space, indicating that the classification of LADS is more strongly influenced by the intrinsic pulsation modes than by non-linear effects.
\end{itemize}

\subsubsection{Discussion}

The clustering analysis shows that the traditional amplitude-based classification of $\delta$ Scuti stars into HADS and LADS partially aligns with the groups identified using frequency-domain features. However, the inclusion of non-linear features reveals additional substructure within both classes, indicating that the HADS/LADS dichotomy does not fully capture the diversity of pulsational behaviour in these stars. Therefore, these findings question the simplicity of the amplitude-based classification and advocate for a more comprehensive approach that incorporates multiple observable variables.

These results provide independent support for the idea that non-linear behaviour in HADS and LADS may arise from different physical regimes. HADS predominantly occupy clusters characterized by simpler non-linear signatures, consistent with non-linearities driven by the interaction between pulsation and the stellar envelope, as expected for predominantly slow rotators. In contrast, the emergence of a cluster dominated by subtraction combinations indicates a distinct non-linear regime. Subtraction terms have been proposed as possible signatures of resonance-related interactions \citep{Lares-Martiz2022}; however, this interpretation remains conjectural and may also reflect methodological effects, such as the choice of parent frequencies.

Accordingly, the subgroups identified within HADS may be physically related to differences in the properties of the stellar envelope, such as its thickness or response to pulsation. Subgrouping in LADS, on the other hand, may reflect different ways in which resonance phenomena can arise, for example through rotation-induced mode coupling, near-resonances between independent modes, or parametric resonances involving p- and g-modes \citep{Bowman2016}. However, all these interpretations should be regarded as working hypotheses rather than as evidence for specific physical mechanisms. Without independent constraints from additional observables—such as effective temperature, surface gravity, metallicity, rotation, binarity, or evolutionary status, it is not possible to uniquely associate the identified clusters with particular physical processes.

Nevertheless, the results demonstrate that non-linear features provide valuable discriminatory power and can reveal structure beyond traditional amplitude-based classifications. Future work will explore the physical origin of these clusters by combining the clustering results with independent observational parameters and detailed pulsation modelling, with the aim of assessing the role of rotation, metallicity, binarity, and non-linear mode coupling in shaping the observed pulsation spectra.



\section{Conclusion and Future Work}


The findings of this study have the potential to advance our understanding of stellar evolution, particularly in the context of intermediate-mass stars. By uncovering new subgroups and refining the classification of $\delta$ Scuti stars, this research could pave the way for more accurate models of stellar interiors and pulsation mechanisms. Furthermore, the application of machine learning techniques to asteroseismology is a promising area for future research offering new tools to analyze the vast amounts of data generated by current and future space missions.

This study has explored the application of machine learning clustering techniques to the classification of $\delta$ Scuti stars, with the goal of uncovering intrinsic subgroups and refining our understanding of pulsation and internal structure of these stars. The findings challenge the traditional amplitude-based classification (HADS/LADS) and highlight the importance of incorporating multiple observable variables, including non-linear features, to capture the complexity of these stars.

The clustering results demonstrate that the traditional amplitude-based classification (HADS/LADS) partially aligns with the clusters identified using frequency-domain features. However, the inclusion of non-linear features reveals additional subgroups that suggest the presence of different pulsation mechanisms and internal structures. These findings have several important implications:

\begin{itemize}
    \item \textbf{Limitations of Amplitude-Based Classification:} The light curve amplitude, although an important variable, is not sufficient to fully capture the complexity of $\delta$ Scuti stars. The amplitude masks the underlying dynamics of oscillations and their non-linear effects, leading to potential misclassifications and ambiguities.
    
    \item \textbf{Importance of Non-Linear Features:} Non-linear features, such as harmonics and combination frequencies, play a crucial role in differentiating subgroups within $\delta$ Scuti stars. These features provide valuable insights into the internal structures and dynamics of these stars, particularly in the context of resonance effects and mode coupling.

\end{itemize}

The study highlights the potential of machine learning techniques to advance our understanding of $\delta$ Scuti stars. By uncovering new subgroups and refining the classification of these stars, this research contributes to the broader field of stellar astrophysics and paves the way for more accurate models of stellar interiors and pulsation mechanisms.

The findings of this study open several avenues for future research, which can be categorized into the following approaches:

\begin{itemize}
    \item \textbf{Massive Sample Clustering Study in LADS:} Given that HADS represent only 0.8\% of $\delta$ Scuti stars, future research should focus on analyzing a larger sample of LADS. A larger dataset would provide more robust clustering results and allow the identification of additional subgroups in LADS.
    
    \item \textbf{Analysis of Cluster Patterns:} A post-study of the identified clusters, incorporating additional data such as spectroscopy, could provide valuable insights into the physical properties and mechanisms associated with each cluster. For example, analyzing the clusters in the Hertzsprung-Russell (HR) diagram or comparing them with stellar evolution models could help validate the clustering results and refine the classification of $\delta$ Scuti stars.
    
    \item \textbf{Iterative Work Between BPM Criteria Development and Cluster Pattern Recognition:} The BPM method, used to process the dataset, could be further refined based on the clustering results. An iterative process, combining clustering analysis with improvements in the BPM criteria, could lead to more accurate identification of parent and child frequencies, enhancing our understanding of non-linear mechanisms in $\delta$ Scuti stars.
    
    \item \textbf{Integration of Additional Features:} Future studies could incorporate additional features, such as metallicity, rotation rates, and binarity, to further refine the clustering results. These features could provide additional dimensions for clustering and help uncover new patterns within $\delta$ Scuti stars.
    
    \item \textbf{Validation with Independent Datasets:} Clustering results might be validated using independent datasets from other space missions, such as the future PLATO space mission \citep{rauer2014plato} . This would help confirm the robustness of the findings and ensure that the clustering patterns are not specific to the current dataset.
    
    \item \textbf{Development of Non-Linear Models:} The study highlights the importance of non-linear effects in $\delta$ Scuti stars. Future research should focus on developing non-linear models that can accurately capture the complex interactions between pulsation modes and the internal structure. These models could provide deeper insights into the selection mechanisms of pulsations and non-linear effects associated with $\delta$ Scuti stars.
\end{itemize}

This study represents a significant step forward in the classification and understanding of $\delta$ Scuti stars. By challenging the traditional amplitude-based classification and incorporating non-linear features, the research provides a more nuanced understanding of these fascinating objects. The findings underscore the importance of considering multiple observable variables and non-linear effects in the study of stellar pulsations, paving the way for future advancements in stellar astrophysics.

The application of machine learning techniques to asteroseismology is a promising avenue for future research, offering new tools to analyze the vast amounts of data generated by current and future space missions. By continuing to refine our understanding of $\delta$ Scuti stars, we can gain valuable insights into stellar evolution, internal structure, and the selection mechanisms of stellar pulsations.

\begin{acknowledgements}
      This manuscript uses data that are publicly available from the Mikulski Archive for Space Telescopes (MAST). JRR and JPG acknowledge financial support from project PID2023-149439NB-C42 from the 'Proyectos de Generación de Conocimiento' and from the Severo Ochoa grant CEX2021-001131-S funded by MICIU/AEI/10.13039/501100011033 and FEDER, EU.
\end{acknowledgements}

\newpage
\bibliographystyle{./sty/astron}
\bibliography{references.bib}

\end{document}